\documentclass[%
reprint,
 amsmath,amssymb,
 aps,
showpacs,showkeys,]{revtex4-1}

\usepackage{graphicx}
\usepackage{dcolumn}
\usepackage{bm}

\begin{document}


\title{Energy-independent optical  $^{1}S_{0}NN$ potential\\
from Marchenko equation}

\author{N. A. Khokhlov}
\email{nikolakhokhlov@yandex.ru}
\affiliation{%
 Southwest State University, Kursk, Russia
}%
\author{L. I. Studenikina}%
\email{sli-kursk@yandex.ru}
\affiliation{%
	MIREA – Russian Technological University, Moscow, Russia
}%
\date{\today}

\begin{abstract}
We present a new algebraic method for solving the inverse problem of quantum scattering theory based on the Marchenko theory. 
We applied a triangular wave set for the Marchenko equation kernel expansion in a separable form. 
The separable form allows a reduction of the Marchenko equation to a system of linear equations. 
For the zero orbital angular momentum, a linear expression of the kernel expansion coefficients is obtained in terms of the Fourier series coefficients of a function depending on the momentum $q$ and determined by the scattering data in the finite range of $q$. 
It is shown that a Fourier series on a finite momentum range  ($0<q<\pi/h$) of a $q(1-S)$ function ($S$ is the scattering matrix) defines the potential function of the corresponding radial Schrödinger equation with $h$-step accuracy. A numerical algorithm is developed for the reconstruction of the optical potential from scattering data.
The developed procedure is applied to analyze the $^{1}S_{0}NN$ data up to 3 GeV. It is shown that these data are described by optical energy-independent partial potential. 
\end{abstract}
\maketitle

\section{\label{sec:intro}INTRODUCTION}
The inverse problem of quantum scattering is essential for various physical applications such as the internuclear potential extraction from scattering data and similar problems. 
The main approaches to solving the problem are Marchenko, Krein, and Gelfand-Levitan theories \cite{Gelfand,Agranovich1963,Marchenko1977,Krein1955,Levitan1984,Newton,Chadan}. 
The ill-posedness of the inverse problem complicates its numerical solution. So far, the development of robust methods for solving the problem remains a fundamental challenge for applications. 
This paper considers a new algebraic method for solving the inverse problem of quantum scattering theory. We derive the method from the Marchenko theory. 
To this end, we propose a  novel numerical solution of the Marchenko equation based on the integral kernel approximation as a separable series in the triangular and rectangular wave sets. Then we show that the series coefficients may be calculated directly from the scattering data on the finite segment of the momentum axis. 
We offer an exact relationship between the potential function accuracy and the range of known scattering data. 

The concept of optical potential (OP) is a useful tool in many branches of nuclear physics. There are a number of microscopically derived \cite{Timofeyuk2020,Burrows2020,Tung2020,Dinmore2019,Vorabbi2018,Wang2018,Atkinson2018,Yahya2018,%
	Lee2018,Vorabbi2017,Fernandez-Soler2017,Guo2017,Rotureau2017,Vorabbi2016} as well as phenomenological \cite{Vorabbi2018,Guo2017,Kobos1985,Hlophe2013,Khokhlov2006,Xu2016,Ivanov2016,Quinonez2020,Jaghoub2018,Titus2016} optical potentials used for description of nuclei and nuclear reactions. Nucleon-nucleon potentials are used as an input for  (semi)microscopic descriptions of nuclei and nuclear reactions \cite{Timofeyuk2020,Burrows2020,Vorabbi2018,Vorabbi2017,Guo2017,Xu2016,Vorabbi2016}. 
We analyzed the Marchenko theory \cite{Agranovich1963,Marchenko1977} and found that it is applicable not only to  unitary $S$-matrices but also to non-unitary $S$-matrices describing absorption. That is, the Marchenko equation and our
algebraic form of the Marchenko equation allow us to reconstruct local and energy-independent OP from an absorbing $S$-matrix and characteristics of corresponding bound states. We applied the developed formalism to analyze the $^{1}S_{0}NN$ data up to 3 GeV and showed that these data are described by energy-independent optical  partial potential. 
Our results contradict conclusions of \cite{Fernandez-Soler2017} where they state that "... the optical potential with a repulsive core exhibits a strong energy dependence whereas the optical potential with the structural core is characterized by a rather adiabatic energy dependence ..." On the contrary, we reconstructed from the scattering data local and energy-independent $NN$ soft core OP.

\section{Marchenko equation in an algebraic form}
We write the radial Schrödinger equation in the form:
\begin{equation}
	\label{f1}
	\left(\frac{d^{2}}{d r^{2}}-\frac{l(l+1)}{r^{2}}-V(r)+q^{2}\right) \psi(r, q)=0.
\end{equation}
Initial data for the Marchenko method [1] are:
\begin{equation}
	\label{f2}
		\left\{S(q),(0<q<\infty), \tilde{q}_{j}, M_{j}, j=1, \ldots, n\right\},
	\end{equation}
where $S(q)=e^{2 \imath \delta(q)}$  is a scattering matrix dependent on the momentum $q$. The $S$-matrix defines asymptotic behavior at  $r \rightarrow+\infty$ of regular at $r=0$  solutions of Eq.~(\ref{f1}) for $q \geq 0 ;\ \tilde{q}_{j}^{2}=E_{j} \leq 0, E_{j}$  is $j$-th bound state energy ($-\imath \tilde{q}_{j} \geq 0$); $M_{j}$  is $j$-th bound state asymptotic constant. The Marchenko equation is a Fredholm integral equation of the second kind:
\begin{equation}
\label{f3}	F(x, y)+L(x, y)+\int_{x}^{+\infty} L(x, t) F(t, y) d t=0
\end{equation}
We write the kernel function as 
\begin{multline}
F(x, y)=\frac{1}{2 \pi} \int_{-\infty}^{+\infty} h_{l}^{+}(q x)[1-S(q)] h_{l}^{+}(q y) d q \\	
+\sum_{j=1}^{n_{b}} h_{l}^{+}\left(\tilde{q}_{j} x\right) M_{j}^{2} h_{l}^{+}\left(\tilde{q}_{j} y\right)\\
=\frac{1}{2 \pi} \int_{-\infty}^{+\infty} h_{l}^{+}(q x) Y(q) h_{l}^{+}(q y) d q
\label{f4}	
\end{multline}
where
\begin{equation}
\label{f5}	Y(q)=\left[1-S(q)-i \sum_{j=1}^{n_{b}} M_{j}^{2}\left(q-\tilde{q}_{j}\right)^{-1}\right]
\end{equation}
Solution  of Eq.~(\ref{f3}) gives the potential of Eq.~(\ref{f1}):
\begin{equation}
	V(r)=-2 \frac{d L(r, r)}{d r} \label{f6}
\end{equation}
There are many computational approaches for the solution of Fredholm integral equations of the second kind. Many of the methods use an equation kernel's series expansion  \cite{eprint7,eprint8,eprint9,eprint10,eprint11,eprint12,eprint13,eprint14}. We also use this technique. Assuming the finite range $R$ of the bounded potential function, we approximate the kernel function as
\begin{equation}
\label{f7}	F(x, y) \approx \sum_{k, j=0}^{N} \Delta_{k}(x) F_{k, j} \Delta_{j}(y)
\end{equation}
where  $F_{k, j} \equiv F(k h, j h)$, and the basis functions are

\begin{equation}
\left. \begin{array}{l}	\Delta_{0}(x)=\left\{\begin{array}{lll}
		0 & \text { for } & |x|>h, \\
		1+x / h & \text { for } & -h \leq x \leq 0, \\
		1-x / h & \text { for } & 0<x \leq h;
	\end{array}\right.\\
\Delta_{n}(x)=\Delta_{0}(x-h n) 
\end{array}\right\} \label{f8}
\end{equation}
where $h$ is some step, and $R = Nh$. 
Decreasing the step $h$, one can approach the kernel arbitrarily close at all points. As a result, the kernel is presented in a separable form. We solve Eq.~(\ref{f3}) substituting
\begin{equation}
\label{f9}	L(x, y) \approx \sum_{j=0}^{N} P_{j}(x) \Delta_{j}(y)
\end{equation}
Substitution of Eqs.~(\ref{f7}) and (\ref{f9}) into Eq.~(\ref{f3}), and taking into account the linear independence of the basis functions, gives 
\begin{widetext}
\begin{equation}
			\sum_{m=0}^{N}\left(\delta_{j\, m}+\sum_{n=0}^{N}\left[\int_{x}^{\infty} \Delta_{m}(t) \Delta_{n}(t) d t\right] F_{n, j}\right) P_{m}(x)	
=-\sum_{k=0}^{N} \Delta_{k}(x) F_{k, j}
	\label{f10}	
\end{equation}
We need values of  $P_{k}(h p) \equiv P_{p, k}$ $(p,k = 0,..,N)$.  In this case integrals in Eq.~(\ref{f10}) may be calculated
\begin{equation}
\zeta_{n\, m\, p}=\int_{p h}^{\infty} \Delta_{m}(t) \Delta_{n}(t) d t	
	=\frac{h}{6}\left(2 \delta_{n\, n}\left(\delta_{n\, p}+2 \eta_{n \geq p+1}\right)
+\delta_{n\,(m-1)} \eta_{n \geq p}+\delta_{n\,(m+1)} \eta_{m \geq p}\right)
	\label{f10b}	
\end{equation}
\end{widetext}
Here, along with the Kronecker symbols  $\delta_{k\, p}$, symbols  $\eta_{a}$ are introduced, which are equal to one if the logical expression $a$ is true, and are equal to zero otherwise. Considering also that  $\Delta_{k}(h p) \equiv \delta_{k\, p}$, we finally get a system of equations 
\begin{equation}
\label{f11}	\sum_{m=0}^{N}\left(\delta_{j\, m}+\sum_{n=0}^{N} \zeta_{n\, m\, p} F_{n, j}\right) P_{p m}=-F_{p, j}
\end{equation}
for each $j,p = 0,..,N$.   
Solution of Eq.~(\ref{f11}) gives  $P_{k}(h p) \equiv P_{p, k}$. Potential values at points $r = hp$ $(p = 0,..,N)$ are determined from Eq.~(\ref{f6}) by some finite difference formula.   

\begin{figure}[htb]
	\centerline{\includegraphics[width=0.45\textwidth]{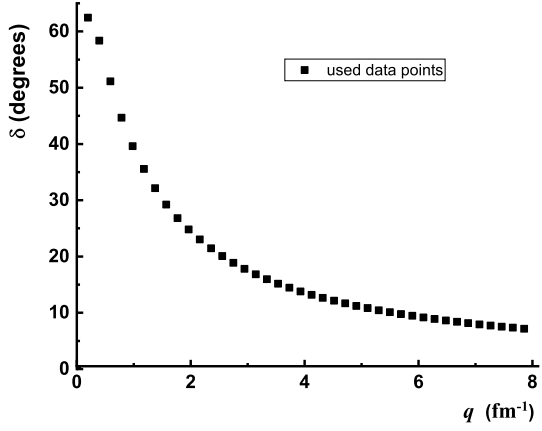}}
	\caption{\label{fig:expphase}Data used to reconstruct $V(r)=V_{0}\exp(-ar)$, where $V_{0}=-3\ fm^{-2}= -124.5\ MeV$, $a=1.5\ fm^{-1}$. Units correspond to the $NN$ system.}
\end{figure}

\begin{figure}[htb]
	\centerline{\includegraphics[width=0.5\textwidth]{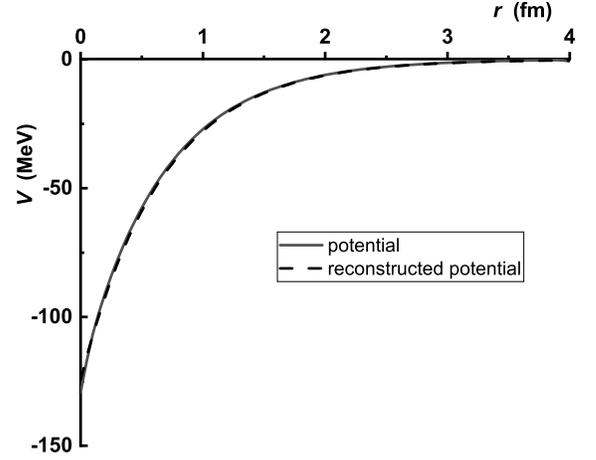}}
	\caption{\label{fig:exppot}Initial and reconstructed potentials:  $V(r)=V_{0}\exp(-ar)$, where $V_{0}=-3\ fm^{-2}= -124.5\ MeV$, $a=1.5\ fm^{-1}$. Units correspond to the $NN$ system.}
\end{figure}

Next, we consider the case $l = 0$, for which $h_{l}^{+}(q x)=e^{\imath q x}$  and
\begin{equation}
\label{f11b}	F(x, y)=F(x+y)=\frac{1}{2 \pi} \int_{-\infty}^{+\infty} e^{\imath q(x+y)} Y(q) d q.
\end{equation}
We approximate the kernel as follows:
\begin{equation}
\label{f12}	F(x, y)=F(x+y) \approx \sum_{k=-2 N}^{2 N} F_{0, k} H_{k}(x+y)
\end{equation}
where $F_{0, k} \equiv F(k h)$  as in Eq.~(\ref{f7}) for $l=0$, and the used basis set is
\begin{equation}
\left. \begin{array}{l} H_{0}(x)=\left\{\begin{array}{lll}
		0 & \text { for } & x<0, \\
		1 & \text { for } & 0 \leq x \leq h, \\
		0 & \text { for } & x>h,
	\end{array}\right.\\
		H_{n}(x)=H_{0}(x-h n).
	\end{array}\right\} \label{f13}
\end{equation}
\begin{widetext}
The Fourier transform of the basis set Eq.~(\ref{f13}) is
\begin{equation}
	\tilde{\mathrm{H}}_{k}(q)=\int_{-\infty}^{\infty} \mathrm{H}_{k}(x) e^{-\imath q x} d x=\frac{\imath\left(e^{-i q h}-1\right)}{q} e^{-\imath q h k}. \label{f13b}
\end{equation}
The function $Y(q)$ may be presented as
\begin{equation}
	Y(q)=\sum_{k=-2 N}^{2 N} F_{0, k} \tilde{\mathrm{H}}_{k}(q)=\sum_{k=-2 N}^{2 N} F_{0, k}  \frac{\imath\left(e^{-\imath q h}-1\right)}{q }  e^{-\imath q h k}. \label{f13c}
\end{equation}
The last relationship may be rearranged
\begin{multline}
q Y(q)= \imath \sum_{k=-2 N}^{2 N} F_{0, k} \left(e^{-\imath q h}-1\right)  e^{-\imath q h k}\\
=i \sum_{k=-2 N+1}^{2 N}\left(F_{0, k-1}-F_{0, k}\right) e^{-\imath q h k}
+\imath\left(-F_{0,-2 N}\right) e^{\imath q h 2 N}+\imath\left(F_{0,2 N}\right) e^{-\imath q h(2 N+1)}.
	\label{f13d}	
\end{multline}
Thus, the left side of the expression is represented as a Fourier series on the interval $-\pi / h \leq q \leq \pi / h$. Taking into account that  $Y(-q)=Y^{*}(q)$, we get
\begin{multline}
	\left.\begin{array}{l}
		-F_{0,-2 N}=\frac{h}{\pi} \int_{0}^{\pi / h}  \text{Im}\left( Y(q) e^{-\imath q h 2 N}\right)q d q; \\
		\\
		F_{0, k-1}-F_{0, k}=\frac{h}{\pi} \int_{0}^{\pi / h}  \text{Im}\left( Y(q) e^{\imath q h  k}\right) q d q 		\text{ for } k=-2 N+1, \ldots, 2 N-1 ; \\
		\\
		F_{0,2 N}=\frac{h}{\pi} \int_{0}^{\pi / h}  \text{Im}\left( Y(q) e^{\imath q h(2 N+1)}\right) q d q.
	\end{array}\right\}\label{f14}
\end{multline}
\end{widetext}
The system~(\ref{f14}) is solved recursively from  $F_{0,2 N}$. Thus, the range of known scattering data defines the step value $h$   and, therefore, the inversion accuracy.
Calculation  results for the potential function  $V(r)=-3 \exp (-3 r / 2)$  are presented in Figs.~\ref{fig:expphase},~\ref{fig:exppot}, where  $h = 0.04$, $R = 4$. $S$-matrix was calculated at points shown in Fig.~\ref{fig:expphase} up to $q = 8$.  The $S$-matrix was interpolated by a quadratic spline in the range $0 < q < 8$. For $q > 8$ the $S$-matrix was approximated as asymptotic $S(q)\approx\exp (-2 i A / q)$  for $q>8$, where $A$ was calculated at $q = 8$. 

\section{Energy-independent optical  $^{1}S_{0}NN$ potential}
Realistic potentials derived unambiguously from inverse theories should describe scattering data from zero to infinite energy. It seems that it is only possible if the available scattering data approach the asymptotic region below the relativistic region. It is unnecessary because relativistic two-particle potential models may be presented in the non-relativistic form \cite{Keister1991}. Another problem is the presence of closed channels whose characteristics are not known. It is usually assumed (for example, for an $NN$ system) that below the inelasticity threshold,  effects of closed channels can be neglected, and a real $NN$ potential may describe the interaction of nucleons. 
This assumption is a consequence of the ingrained misconception that a complex potential corresponds to a non-unitary matrix.  One can only assert that the $S$-matrix is unitary for a real potential. 

We have carefully analyzed the Marchenko theory \cite{Agranovich1963,Marchenko1977} and found that it applies not only to unitary $S$-matrices but also to non-unitary $S$-matrices describing absorption. That is, the Marchenko theory Eqs.~(\ref{f1}-\ref{f6}) and our algebraic form of the Marchenko equation Eqs. (\ref{f7}-\ref{f13d}) allow to reconstruct local, and energy-independent OP from an absorbing $S$-matrix and corresponding bound states' characteristics.
We present an absorptive single partial channel $S$-matrix on the $q$-axis  as
\begin{equation}
 S(q)=\left\{\begin{array}{lll}
			S_{u}(q)+S_{n}(q) & \text { for } & q>0, \\
			S^{+}_{u}(-q)-S^{+}_{n}(-q) & \text { for } & q<0, 
		\end{array}\right.
		\label{OP_Smatrix}
\end{equation}
where superscript $+$ means hermitian conjugation. For $q>0$ we define
 \begin{multline}
 	S_{u}(q) = e^{2\imath \delta(q)},\\
 		S_{n}(q) = -\sin^2(\rho(q))e^{2\imath \delta(q)},
 \end{multline}
where $\delta(q)$ and $\rho(q)$ are phase shift and inelastisity parameter correspondingly. 
In this case we have instead of Eqs.~(\ref{f14}) the following system
\begin{widetext}
\begin{multline}
	\left.\begin{array}{l}
		-F_{0,-2 N}=\frac{h}{\pi} \int_{0}^{\pi / h} q \left[\text{Im}\left(Y_{u}(q) e^{-\imath q h 2 N}\right)-\imath\text{Re}\left(S_{n}(q) e^{-\imath q h 2 N}\right) \right] d q; \\
		\\
		F_{0, k-1}-F_{0, k}=\frac{h}{\pi} \int_{0}^{\pi / h} q \left[ \text{Im}\left( Y_{u}(q) e^{\imath q h k}\right)-\imath\text{Re}\left(S_{n}(q) e^{\imath q h k}\right)\right] d q 		\text{ for } k=-2 N+1, \ldots, 2 N-1 ; \\
				\\
		F_{0,2 N}=\frac{h}{\pi} \int_{0}^{\pi / h} q \left[ \text{Im}\left( Y_{u}(q) e^{\imath q h(2 N+1)}\right) 
		-\imath\text{Re}\left(S_{n}(q) e^{\imath q h (2 N+1)}\right) \right] d q,
	\end{array}\right\}\label{f14b}
\end{multline}
\end{widetext}
where 
\begin{equation}
	\label{f5f}	Y_{u}(q)=\left[1-S_{u}(q)-i \sum_{j=1}^{n_{b}} M_{j}^{2}\left(q-\tilde{q}_{j}\right)^{-1}\right].
\end{equation}

\section{Results and Conclusions}
We applied the developed formalism to analyze the $^{1}S_{0}$ $NN$ data.
As input data for the reconstruction, we used modern
phase shift analysis data (single-energy solutions) up to
3 GeV \cite{DataScat,cite}. We smoothed  phase shift and inelasticity parameter data for $q>3$~fm$^{-1}$ by the following functions:
\begin{multline}
	\delta(q) \sim - 54.56822/q^{3}+57.55296/q^{2}-15.36687/q,\\
	\rho(q) \sim 101.89881/q^{3}-80.13493/q^{2}+15.88984/q,
	\label{asymp}
\end{multline}
where we fitted the coefficients by the least-squares method. Asymptotics (\ref{asymp}) were used to calculate coefficients of Eqs.~(\ref{f14}) with $h=0.0125$~fm corresponding to  $q_{max}\approx 251.3$~fm$^{-1}$.

\begin{figure}[htb]
	\centerline{\includegraphics[width=0.45\textwidth]{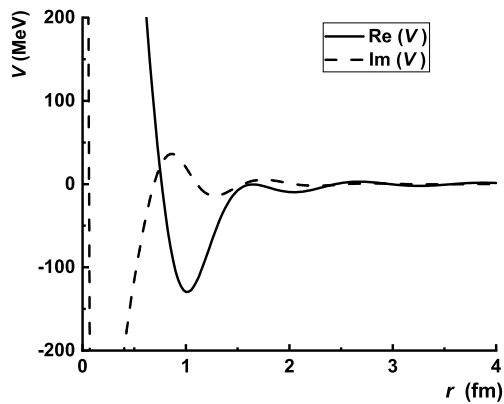}}
	\caption{\label{fig:1S0data}Data used to reconstruct $^{1}S_{0}NN$ potential.}
\end{figure}

\begin{figure}[htb]
	\centerline{\includegraphics[width=0.5\textwidth]{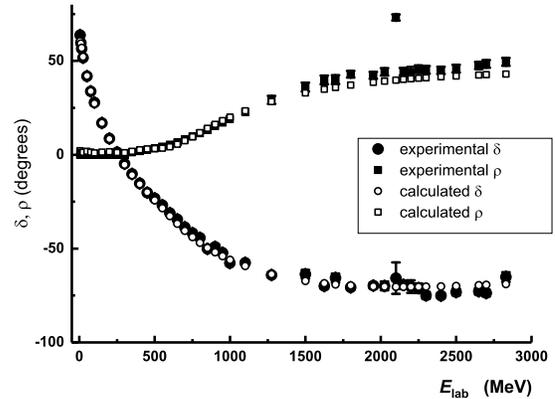}}
	\caption{\label{fig:1S0pot}Reconstructed $^{1}S_{0}NN$ potential.}
\end{figure}

Results of our calculations  show   that these data are described by  energy-independent optical partial potential (Figs.~\ref{fig:1S0data},\ref{fig:1S0pot}). 


 Thus, we presented a solution of the quantum scattering inverse problem for the zero orbital angular momentum, the algorithm of which is as follows. We set the step value $h$, which determines a required accuracy of the potential. From the scattering data, we determine $F_{0,k}$  from Eqs.~(\ref{f14}) for unitary $S$-matrix or from  Eqs.~(\ref{f14b}) for non-unitary $S$-matrix. Solution of Eqs.~(\ref{f11}) gives values of $P_{k}(hp)$ ($p = 0,..,N$). Further, the values of the potential function (\ref{f6}) are determined by some finite difference formula.
Expressions (\ref{f7}-\ref{f12}) give a method for the Marchenko equation's numerical solution for an arbitrary orbital angular momentum $l$, and may be generalized for a case of coupled channels.

Our results contradict conclusions of \cite{Fernandez-Soler2017} claiming that  an OP with a repulsive core exhibits a strong energy dependence. On the contrary, we reconstructed local and energy-independent $NN$ soft core OP with $Re(V(0))\approx 14\ GeV$ and $Im(V(0))\approx 19\ GeV$ .

It may be that some local OPs lead to unsatisfactory description of nuclear reactions \cite{Titus2016} ($(d, p)$ transfer reactions).   Our approach assumes inverse scattering reconstruction of local and energy-independent OP describing all two-particle scattering data (including high energy asymptotics) and bound states. 
Such OPs have not been used in nuclear calculations,  though they may give an adequate description of the off-shell behaviour of the nucleon-nucleon interaction.

The reconstructed $1S0$ $NN$ optical potential may be requested from the author in the Fortran code.

\end{document}